\documentclass[aps]{revtex4}
\usepackage{amsfonts}
\usepackage{amsmath}
\usepackage{amssymb,epsf}
\usepackage{graphicx}

\begin{document}

\title{Black holes in massive gravity as heat engines}
\author{S. H. Hendi$^{1,2}$\footnote{
email address: hendi@shirazu.ac.ir}, B. Eslam
Panah$^{1,2,3}$\footnote{ email address:
behzad.eslampanah@gmail.com}, S. Panahiyan$^{1,4,5}$\footnote{
email address: shahram.panahiyan@uni-jena.de} , H. Liu$^{6}$\footnote{%
email address: hangliu@mail.nankai.edu.cn} and X. -H. Meng$^{6,7}$\footnote{%
email address: xhm@nankai.edu.cn}}
\affiliation{$^1$ Physics Department and Biruni Observatory, College of Sciences, Shiraz
University, Shiraz 71454, Iran\\
$^2$ Research Institute for Astronomy and Astrophysics of Maragha (RIAAM),
P.O. Box 55134-441, Maragha, Iran\\
$^3$ ICRANet, Piazza della Repubblica 10, I-65122 Pescara, Italy\\
$^4$ Helmholtz-Institut Jena, Fr\"{o}belstieg 3, Jena 07743, Germany \\
$^5$ Physics Department, Shahid Beheshti University, Tehran 19839, Iran\\
$^6$ School of Physics, Nankai University, Tianjin 300071, China\\
$^7$ State Key Laboratory of Theoretical Physics, Institute of Theoretical
Physics, Chinese Academy Of Science, Beijing 100190, China}

\begin{abstract}
The paper at hand studies the heat engine provided by black holes in the
presence of massive gravity. The main motivation is to investigate the
effects of massive gravity on different properties of the heat engine. It
will be shown that massive gravity parameters and graviton's mass modify the
efficiency of engine on a significant level. Furthermore, it will be shown
that it is possible to have the heat engine for non-spherical black holes in
massive gravity and we study the effects of topological factor on properties
of the heat engine. Surprisingly, it will be shown that the highest
efficiency for the heat engine belongs to black holes with hyperbolic
horizon, while the lowest one belongs to spherical black holes.
\end{abstract}

\maketitle

\section{Introduction}

Einstein theory of gravity is one of the best theories which has introduced
until now. This theory has some interesting predictions, such as the
existence of gravitational waves which observed by advanced LIGO in $2016$.
But there are some phenomena in which such theory can not explain them,
precisely. For example we refer the reader to the current acceleration of
the universe and the cosmological constant problem. In addition, this theory
predicts the existence of massless spin-$2$ gravitons in which they have two
degrees of freedom. However, there have been some arguments regarding the
possibility of the existence of massive spin-$2$ gravitons, such as the
hierarchy problem and also brane-world gravity solutions (see Refs. \cite%
{MassiveIb,MassiveIc} for more details). These show that, despite its
correctness, the general relativity is not the final theory of gravitation.
On the other hand, the massive gravity includes some interesting properties.
One of them is that this theory could explain the accelerated expansion of
the universe without considering the dark energy. Also, the graviton behaves
like a lattice excitation and exhibits a Drude peak in this theory of
gravity. It is notable that, current experimental data from the observation
of gravitational waves by advanced LIGO requires the graviton mass to be
smaller than the inverse period of orbital motion of the binary system, that
is $m=1.2\times 10^{-22}ev/c^{2}$ \cite{Abbott}.

Another important reason in order to consider the massive gravity is related
to the fact that possibility of the massive graviton help us to understand
the quantum gravity effects \cite{MassiveIa,MassiveId}. With this goal in
mind, Fierz and Pauli introduced a massive theory of gravity in a flat
background \cite{Fierz1939}. The problem with this theory was the fact that
it suffers from vDVZ (van Dam-Veltman-Zakharov) discontinuity. To resolve
this problem, Vainshtein introduced his well known mechanism requiring the
theory being considered in a nonlinear framework. Although Vainshtein
mechanism was a solution to vDVZ discontinuity, it reveals yet another
profound problem of the Fierz and Pauli theory known as Boulware-Deser ghost
\cite{BDghost} which signals instability in the theory of interest. In order
to avoid such instability, several models of massive theory are introduced
by some authors. For example, Bergshoeff, Hohm and Townsend proposed one of
the ghost-free massive theories in three dimensional spacetime which is
known as new massive gravity (NMG) \cite{Newmasssive}. This theory of
massive gravity has been investigated in many literatures \cite%
{NewM1,NewM2,NewM3,NewM4,NewM5}, however, this theory has some problems in
higher dimensions. Another interesting class of massive gravity was
introduced by de Rham, Gabadadze and Tolley (dRGT) \cite{dRGTI,dRGTII}. This
theory is valid in higher dimensions as well. It is notable that the mass
terms in dRGT theory are produced by consideration of a reference metric.
The stability of this massive theory was studied and it was shown that this
theory enjoys absence of the Boulware-Deser ghost \cite{HassanI,HassanII}.
Black hole solutions and their thermodynamical properties with considering
dRGT massive gravity have been investigated in Refs. \cite%
{BHMassiveI,BHMassiveII,BHMassiveIII,BHMassiveIV}. From the perspective of
astrophysics, Katsuragawa et al in Ref. \cite{Katsuragawa}, studied the
neutron star in the context of this theory and showed that, the massive
gravity leads to small deviation from the general relativity (GR). In the
cosmological context, phantom crossing and quintessence limit \cite%
{Saridakis}, bounce and cyclic cosmology \cite{YFCai}, cosmological behavior
\cite{Leon}, and other properties of this gravity have been studied by some
authors \cite{Hinterbichler,Fasiello,Bamba}.

Modification in the reference metric in dRGT theory provides the possibility
of introduction of different classes of dRGT like massive theories. Among
them, one can point out the one introduced by Vegh which has applications in
gauge/gravity duality \cite{Vegh}. This theory is similar to dRGT theory
with a difference that its reference metric is a singular one. Considering
this theory of massive gravity, Vegh in Ref. \cite{Vegh}, showed that
graviton may behave like a lattice and exhibits a Drude peak. Also, it was
pointed out that for arbitrary singular metric, this theory of massive
gravity is ghost-free and stable \cite{HZhang}. Using this massive theory of
gravity, different classes of the charged black hole solutions have been
studied in Refs. \cite{Cai2015,VeMassI,VeMassII,VeMassIII}. In addition, the
existence of van der Waals like behavior in extended phase space has been
investigated in Refs. \cite{PVMassI,PVMassII,PVMassIII,PVMassIV,PVMassV}.
Holographic conductivity in this gravity has been explored in Refs. \cite%
{Alberte,Zhou,Dehyadegari}. Moreover, magnetic solutions of such theory have
been addressed in Ref. \cite{Magmass}. From the astrophysical point of view,
the hydrostatic equilibrium equation of neutron stars by considering this
theory of massive gravity was obtained in Ref. \cite{NeutronMass}, and it
was shown that the maximum mass of neutron stars can be about $3.8M_{\odot }$
(where $M_{\odot }$ is mass of the Sun).

Among the other achievements of the massive theory of gravity, one can point
out the following ones: (i) the cosmological constant could be realized by
massive terms without the need of introduction of cosmological constant into
the action \cite{Gumrukcuoglu,Gratia,Kobayash}. (ii) addressing the
acceleration expansion of the universe without cosmological constant and
through the properties of massive gravity \cite{DeffayetI,DeffayetII}. (iii)
in large scale, the effects of massive gravity could allow the universe to
accelerate while in the small scale, the effects are not on significant
level and GR is dominant. This provides a better coincidence with
experimental observations. \cite{DvaliI,DvaliII}. (v) the massive gravity
provides additional polarization for gravitational waves which modifies its
speed of propagation \cite{Will}. This indicates that there will be
modification in the production of gravitational waves during inflation as
well \cite{Mohseni,GumrukcuogluII}. (vi) the maximum mass of neutron stars
in massive gravity can be more than $3.2M_{\odot }$ \cite{NeutronMass} ($%
3.2M_{\odot }$ is the maximum mass of a neutron star in GR \cite{Ruffini}).

Here, we consider Vegh's approach from the massive gravity. The action of $d$%
-dimensional Einstein-massive gravity with negative cosmological constant in
the presence Maxwell source is
\begin{equation}
I=-\frac{1}{16\pi }\int d^{d}x\sqrt{-g}\left[ R-2\Lambda +L\left( F\right)
+m^{2}\sum_{i}^{4}c_{i}U_{i}\left( g,f\right) \right] ,  \label{Action}
\end{equation}%
where $\Lambda =-\frac{\left( d-1\right) \left( d-2\right) }{2l^{2}}$ is the
negative cosmological constant, $R$ is the scalar curvature and $f$ is a
fixed symmetric tensor. In Eq. (\ref{Action}), $c_{i}$ are constants and $%
U_{i}$ are symmetric polynomials of the eigenvalues of the $d\times d$
matrix $K_{\nu }^{\mu }=\sqrt{g^{\mu \alpha }f_{\alpha \nu }}$ which can be
written as
\begin{eqnarray}
U_{1} &=&\left[ K\right] ,\ \ \ \ U_{2}=\left[ K\right] ^{2}-\left[ K^{2}%
\right] ,\ \ \ \ U_{3}=\left[ K\right] ^{3}-3\left[ K\right] \left[ K^{2}%
\right] +2\left[ K^{3}\right]   \notag \\
&&  \notag \\
U_{4} &=&\left[ K\right] ^{4}-6\left[ K^{2}\right] \left[ K\right] ^{2}+8%
\left[ K^{3}\right] \left[ K\right] +3\left[ K^{2}\right] ^{2}-6\left[ K^{4}%
\right] .
\end{eqnarray}

Here, we want to study Maxwell electromagnetic, so the function $L\left(
F\right) $ is
\begin{equation}
L\left( F\right) =-F,
\end{equation}%
where $F=F_{\mu \nu }F^{\mu \nu }$ (in which $F_{\mu \nu }=\partial _{\mu
}A_{\nu }-\partial _{\nu }A_{\mu }$) is the electromagnetic field tensor.
Also, $A_{\mu }$ is the gauge potential. Variation of the action (\ref%
{Action}) with respect to the metric tensor ($g_{\mu \nu }$) and the
electromagnetic field tensor ($F_{\mu \nu }$), lead to
\begin{eqnarray}
G_{\mu \nu }+\Lambda g_{\mu \nu }+\frac{1}{2}g_{\mu \nu }F-2F_{\mu \lambda
}F_{\nu }^{\lambda }+m^{2}X_{\mu \nu } &=&0, \\
&&  \notag \\
\partial _{\mu }\left( \sqrt{-g}F^{\mu \nu }\right) &=&0,  \label{Maxwell eq}
\end{eqnarray}%
where $G_{\mu \nu }$ is the Einstein tensor and $X_{\mu \nu }$ is the
massive term with the following form
\begin{eqnarray}
X_{\mu \nu } &=&-\frac{c_{1}}{2}\left( U_{1}g_{\mu \nu }-K_{\mu \nu }\right)
-\frac{c_{2}}{2}\left( U_{2}g_{\mu \nu }-2U_{1}K_{\mu \nu }+2K_{\mu \nu
}^{2}\right) -\frac{c_{3}}{2}\left( U_{3}g_{\mu \nu }-3U_{2}K_{\mu \nu
}+6U_{1}K_{\mu \nu }^{2}-6K_{\mu \nu }^{3}\right)  \notag \\
&&  \notag \\
&&-\frac{c_{4}}{2}\left( U_{4}g_{\mu \nu }-4U_{3}K_{\mu \nu }+12U_{2}K_{\mu
\nu }^{2}-24U_{1}K_{\mu \nu }^{3}+24K_{\mu \nu }^{4}\right) .
\end{eqnarray}

Black hole thermodynamics has been studied widely and intensively for a long
time ever since the seminal work done by Hawking et al. \cite%
{Hawking1,Hawking2}. The amazing discovery of the thermodynamical property
of black holes helps us to have a deeper understanding of gravity and
realize that the gravitational systems have some profound relations to
thermodynamical systems. The concept of extended phase space was proposed
\cite{Man} by regarding the cosmological constant $\Lambda $ as
thermodynamic pressure $P$ and its conjugate quantity as thermodynamic
volume $V$, as
\begin{equation}
M=H=U+PV,\quad P=-\frac{\Lambda }{8\pi }=\frac{3}{8\pi \ell ^{2}},\quad V=%
\frac{\partial M}{\partial P}\Big |_{S,Q},\quad \Phi =\frac{\partial M}{%
\partial Q}\Big |_{S,P},\quad S=\int_{0}^{r_{+}}T^{-1}\left( \frac{\partial M%
}{\partial r}\right) _{Q,P}dr,
\end{equation}%
where $Q$ is the electric charge and $\Phi $ is its conjugate electric
potential, $S$ and $T$ sand for the entropy and Hawking temperature on event
horizon $r_{+}$, respectively. The new first law of black hole
thermodynamics in the extended phase space is written as
\begin{equation}
dM=TdS+VdP+\Phi dQ+\Omega dJ,  \label{FL}
\end{equation}%
where the black hole mass $M$ should be interpreted as enthalpy rather than
internal energy of the gravitational system, which is indicated by the first
law (\ref{FL}). For the black hole in $d=n+2$ dimensional massive gravity,
the first law reads
\begin{equation}
\begin{split}
dM=& TdS+VdP+\Phi dQ+\frac{V_{n}cm^{2}r_{+}^{n}}{16\pi }dc_{1}+\frac{%
nV_{n}c^{2}m^{2}r_{+}^{n-1}}{16\pi }dc_{2} \\
& +\frac{n(n-1)V_{n}c^{3}m^{2}r_{+}^{n-2}}{16\pi }dc_{3}+\frac{%
n(n-1)(n-2)V_{n}c^{4}m^{2}r_{+}^{n-3}}{16\pi }dc_{4},
\end{split}%
\end{equation}%
where $V_{n}$ is the volume of space spanned by space coordinates and we
have viewed the coupling constants $c_{i}$ as thermodynamic variables. By
employing scaling method, the Smarr relation can be obtained as
\begin{equation}
(n-1)M=nTS-2PV+(n-1)\Phi Q-\frac{V_{n}cc_{1}m^{2}}{16\pi }r_{+}^{n}+\frac{%
n(n-1)V_{n}c^{3}c_{3}m^{2}}{16\pi }r_{+}^{n-2}+\frac{%
n(n-1)(n-2)V_{n}c^{4}c_{4}m^{2}}{8\pi }r_{+}^{n-3}
\end{equation}

Inspired by the first law of black hole thermodynamics,one natural idea is
to introduce the concept of traditional heat engines into the black hole
thermodynamics with both the thermodynamic pressure and volume already
defined in extended phase space \cite%
{Johnson,Sadeghi,JohnsonI,JohnsonII,Bhamidipati,Mo,Setare}. Treating AdS
black holes as heat engines are thought as a possible way that the useful
mechanical work of both static and stationary AdS black holes is allowed to
be extracted from heat energy. While the Penrose Process, known as way to
extract black hole energy, can only be used for rotating black holes, but
Penrose Process can be exerted to black holes in both asymptotically AdS and
flat spacetime. What's more, the black hole energy can also be released and
spread in spacetime in form of gravitational waves by the collision of two
black holes, which has been detected recently \cite{Abbott}.

\section{Thermodynamic Cycle and Heat Engines}

The entropy of these black holes is obtained through the area law \cite%
{EntropyI,EntropyII} in the following form
\begin{equation}
S=\pi r_{+}^{2}.  \label{TotalS}
\end{equation}

Using the new interpretation of the cosmological constant as thermodynamical
pressure ($P\propto \Lambda $) \cite{Man,PVI,PVII,PVIV,PVV,PVVI}, one can
replace the cosmological constant with following relation%
\begin{equation}
P=-\frac{\Lambda }{8\pi },  \label{Lambda}
\end{equation}%
in which, the thermodynamic volume, $V$ (conjugating to pressure),\ is given
by%
\begin{equation}
V=\frac{4\pi }{3}r_{+}^{3}.  \label{V}
\end{equation}

\begin{figure}[t]
\centering
$%
\begin{array}{c}
\epsfxsize=15cm \epsffile{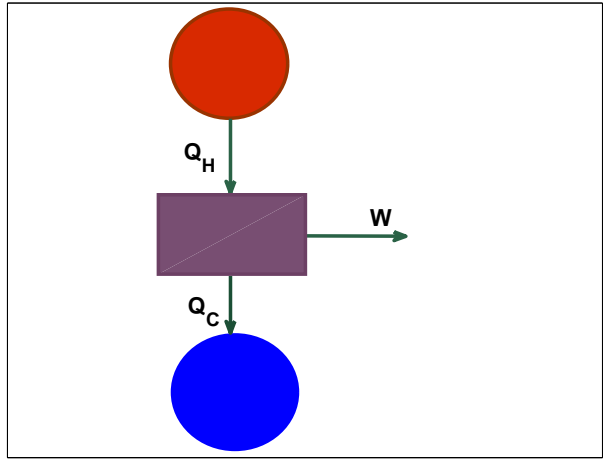}%
\end{array}
$%
\caption{The heat engine flows.}
\label{Fig1}
\end{figure}

Here, we are interested in classical heat engine. It is notable that a heat
engine is a physical system that takes heat from warm reservoir and turns a
part of it into the work while the remaining is dedicated to cold reservoir
(see Fig. (\ref{Fig1})). In order to calculate work done by the heat engine
and given the equation of state, one can use the $P-V$ diagram specified the
heat engine which forms a closed path. For a thermodynamics cycle, one may
extract mechanical work via the $PdV$\ term in the first law of
thermodynamics as $W=Q_{H}-Q_{C}$ (for a thermodynamic cycle, the internal
energy changes is zero ($\triangle U=0$), so the first law of thermodynamics
$\triangle U=\triangle Q-W$ \ reduces to $W=\triangle Q=Q_{H}-Q_{C}$, where $%
Q_{H}$\ is a net input heat flow, $Q_{C}$\ is a net output flow and $W$ is a
net output work), and so we have
\begin{equation}
Q_{H}=W+Q_{C},
\end{equation}%
where the efficiency of heat engine is defined as%
\begin{equation}
\eta =\frac{W}{Q_{H}}=1-\frac{Q_{C}}{Q_{H}},
\end{equation}


It is known that the heat engine depends on the choice of path in the $P-V$
diagram and possibly the equation of state of the black hole in question. It
is notable that, some of the classical cycles involve a pair of isotherms at
temperatures $T_{H}$ and $T_{C}$, in which $T_{H}>T_{C}$. For example, in
the Carnot cycle, there is a pair of isotherms with different temperatures
in which this cycle has maximum efficiency and it is described as $\eta =1-%
\frac{T_{C}}{T_{H}}$. In this cycle, there is an isothermal expansion when
the system absorbs some heat and an isothermal compression during expulsion
of some heat of the system. Using different methods, one can connect these
two systems to each other. The first method is isochoric path, like
classical Stirling cycle, and the second one is adiabatic path, like
classical Carnot cycle. Therefore, the form of path for the definition of
cycle is important. As we know, for the static black holes, the entropy $S$
and the thermodynamic volume $V$\ are related by Eqs. (\ref{TotalS}) and (%
\ref{V}) as $S=\pi r_{+}^{2}=\pi \left( \frac{3V}{4\pi }\right) ^{2/3}$.%
\textbf{\ }It means that adiabatic and isochores are the same, so Carnot and
Stirling methods coincide with each other. Therefore, the efficiency of
cycle can be calculated easily.

\begin{figure}[t]
\centering
$%
\begin{array}{c}
\epsfxsize=15cm \epsffile{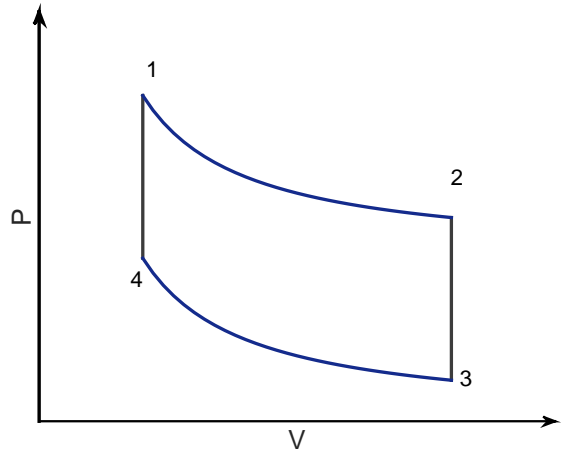}%
\end{array}
$%
\caption{Carnot cycle.}
\label{Fig2}
\end{figure}

So along the upper isotherm (Fig. (\ref{Fig2})) and by using of Eqs. (\ref%
{TotalS}) and (\ref{V}), we have the following heat flow
\begin{equation}
Q_{H}=T_{H}\nabla S_{1\rightarrow 2}=\pi T_{H}\left( \frac{3}{4\pi }\right)
^{2/3}\left( V_{2}^{2/3}-V_{1}^{2/3}\right) ,
\end{equation}%
and also along the lower isotherm (Fig. (\ref{Fig2})) and by using Eqs. (\ref%
{TotalS}) and (\ref{V}), isotherm the heat flow will be
\begin{equation}
Q_{C}=T_{C}\nabla S_{3\rightarrow 4}=\pi T_{C}\left( \frac{3}{4\pi }\right)
^{2/3}\left( V_{3}^{2/3}-V_{4}^{2/3}\right) ,
\end{equation}%
which according to Fig. (\ref{Fig2}), we have $V_{1}=V_{4}$ and $V_{2}=V_{3}$%
, so the efficiency becomes
\begin{equation}
\eta =1-\frac{Q_{C}}{Q_{H}}=1-\frac{T_{C}}{T_{H}}.
\end{equation}

The heat engine for black holes was proposed by Johnson in 2014 \cite%
{Johnson}. Using the concepts introduced by Johnson, the heat engines
provided by other types of black holes have been investigated. For example;
the heat engine for Kerr AdS and dyonic black holes \cite{Sadeghi},
Gauss-Bonnet \cite{JohnsonI}, Born-Infeld AdS \cite{JohnsonII}, dilatonic
Born-Infeld \cite{Bhamidipati}, BTZ \cite{Mo} and polytropic black holes
\cite{Setare} have been studied.

\section{Charged black holes in massive gravity as Heat Engines}

In this section, the $4$-dimensional static charged black holes in the
context of massive gravity with adS asymptotes are introduced. For this
purpose, we consider a metric of $4$-dimensional spacetime in the following
form
\begin{equation}
ds^{2}=-f(r)dt^{2}+f^{-1}(r)dr^{2}+r^{2}\left( d\theta ^{2}+\sin ^{2}\theta
d\varphi ^{2}\right) ,  \label{Metric}
\end{equation}%
with the following reference metric \cite{Cai2015,VeMassI}
\begin{equation}
f_{\mu \nu }=diag(0,0,c^{2},c^{2}\sin ^{2}\theta ),  \label{f11}
\end{equation}%
where in the above equation, $c$ is a positive constant. Using the reference
metric introduced in Eq. (\ref{f11}) for $4$-dimensional spacetime, $%
\mathcal{U}_{i}$'s are in the following forms \cite{Cai2015,VeMassI}
\begin{equation}
\mathcal{U}_{1}=\frac{2c}{r},\text{ \ \ }\mathcal{U}_{2}=\frac{2c^{2}}{r^{2}}%
,\text{ \ }\mathcal{U}_{3}=0,\text{ \ }\mathcal{U}_{4}=0.  \notag
\end{equation}

Using the gauge potential ansatz $A_{\mu }=h(r)\delta _{\mu }^{0}$ in
electromagnetic equation (\ref{Maxwell eq}) and considering the metric (\ref%
{Metric}), the metric function $f(r)$ is obtained in Refs. \cite%
{Cai2015,VeMassI} as
\begin{equation}
f\left( r\right) =1-\frac{m_{0}}{r}-\frac{\Lambda }{3}r^{2}+\frac{q^{2}}{%
r^{2}}+m^{2}\left( \frac{cc_{1}}{2}r+c^{2}c_{2}\right) ,  \label{f(r)}
\end{equation}%
in which $q$ and $m_{0}$ are integration constants related to the electrical
charge and the total mass of black holes, respectively.

The temperature of these black holes could be obtain by employing definition
of Hawking temperature which is based on the surface gravity on the outer
horizon, $r_{+}$, and by considering Eq. (\ref{f(r)}), it will be
\begin{equation}
T=\frac{1}{4\pi r_{+}}\left[ 1-\Lambda r_{+}^{2}-\frac{q^{2}}{r_{+}^{2}}%
+m^{2}cc_{1}r_{+}+m^{2}c^{2}c_{2}\right] ,  \label{Tem}
\end{equation}%
which by using Eqs. (\ref{TotalS})-(\ref{V}), we can rewrite the temperature
in terms of $S$ and $P$ as
\begin{equation}
T=\frac{1}{4\sqrt{\pi S}}\left[ 1+8SP-\frac{\pi q^{2}}{S}+m^{2}cc_{1}\sqrt{%
\frac{S}{\pi }}+m^{2}c^{2}c_{2}\right] .  \label{T1}
\end{equation}

Considering the Eq. (\ref{T1}), one can write the pressure as a function of $%
T$ and $S$ in the following form
\begin{equation}
P=\frac{\sqrt{\pi }T}{2\sqrt{S}}-\frac{m^{2}c^{2}c_{2}+1}{8S}+\frac{\pi q^{2}%
}{8S^{2}}-\frac{m^{2}cc_{1}}{8\sqrt{\pi S}}.  \label{P}
\end{equation}

On the other hand, there are two different heat capacities for a system, the
heat capacity at constant pressure and the heat capacity at constant volume.
The heat capacity can be calculated by the standard thermodynamic relations,
as
\begin{equation}
C_{V}=T\frac{\partial S}{\partial T}\left\vert _{_{V}}\right.
~~~~~\&~~~~~C_{P}=T\frac{\partial S}{\partial T}\left\vert _{_{P}}\right. .
\end{equation}

Considering the fact that entropy is a regular function of the thermodynamic
volume $V$ ($S=\pi r_{+}^{2}=\pi \left( \frac{3V}{4\pi }\right) ^{2/3}$),
the heat capacity at constant volume will vanish, $C_{V}=0$. The heat
capacity at constant pressure is calculated as
\begin{equation}
C_{P}=\frac{T}{\frac{\partial T}{\partial S}\left\vert _{_{P}}\right. }=%
\frac{2S\left[ 8S^{2}P+\left( m^{2}c^{2}c_{2}+1\right) S-\pi q^{2}+\frac{%
m^{2}cc_{1}}{\sqrt{\pi }}S^{3/2}\right] }{8S^{2}P-\left(
m^{2}c^{2}c_{2}+1\right) S+3\pi q^{2}}.  \label{Cp}
\end{equation}

Evidently, due to presence of the massive gravity's parameter, obtained $%
C_{P}$ differs from the usual Reissner-Nordstr\"{o}m one. Using the relation
between $S$ and $V$, we can rewrite the equation (\ref{P}) versus $T$ and $V$
as
\begin{equation}
P=\frac{\left( T-\frac{m^{2}cc_{1}}{4\pi }\right) }{V^{1/3}}\left( \frac{\pi
}{6}\right) ^{1/3}-\frac{\left( m^{2}c^{2}c_{2}+1\right) }{2\pi V^{2/3}}%
\left( \frac{\pi }{6}\right) ^{2/3}+\frac{2q^{2}}{\pi V^{4/3}}\left( \frac{%
\pi }{6}\right) ^{4/3}.  \label{PV}
\end{equation}

It is notable that, in the absence of massive gravity ($m^{2}=0$), the above
equation reduces to the following form \cite{Johnson}
\begin{equation}
P=\frac{T}{V^{1/3}}\left( \frac{\pi }{6}\right) ^{1/3}-\frac{1}{2\pi V^{2/3}}%
\left( \frac{\pi }{6}\right) ^{2/3}+\frac{2q^{2}}{\pi V^{4/3}}\left( \frac{%
\pi }{6}\right) ^{4/3}.
\end{equation}

\begin{figure}[t]
\centering
$%
\begin{array}{cc}
\epsfxsize=7cm \epsffile{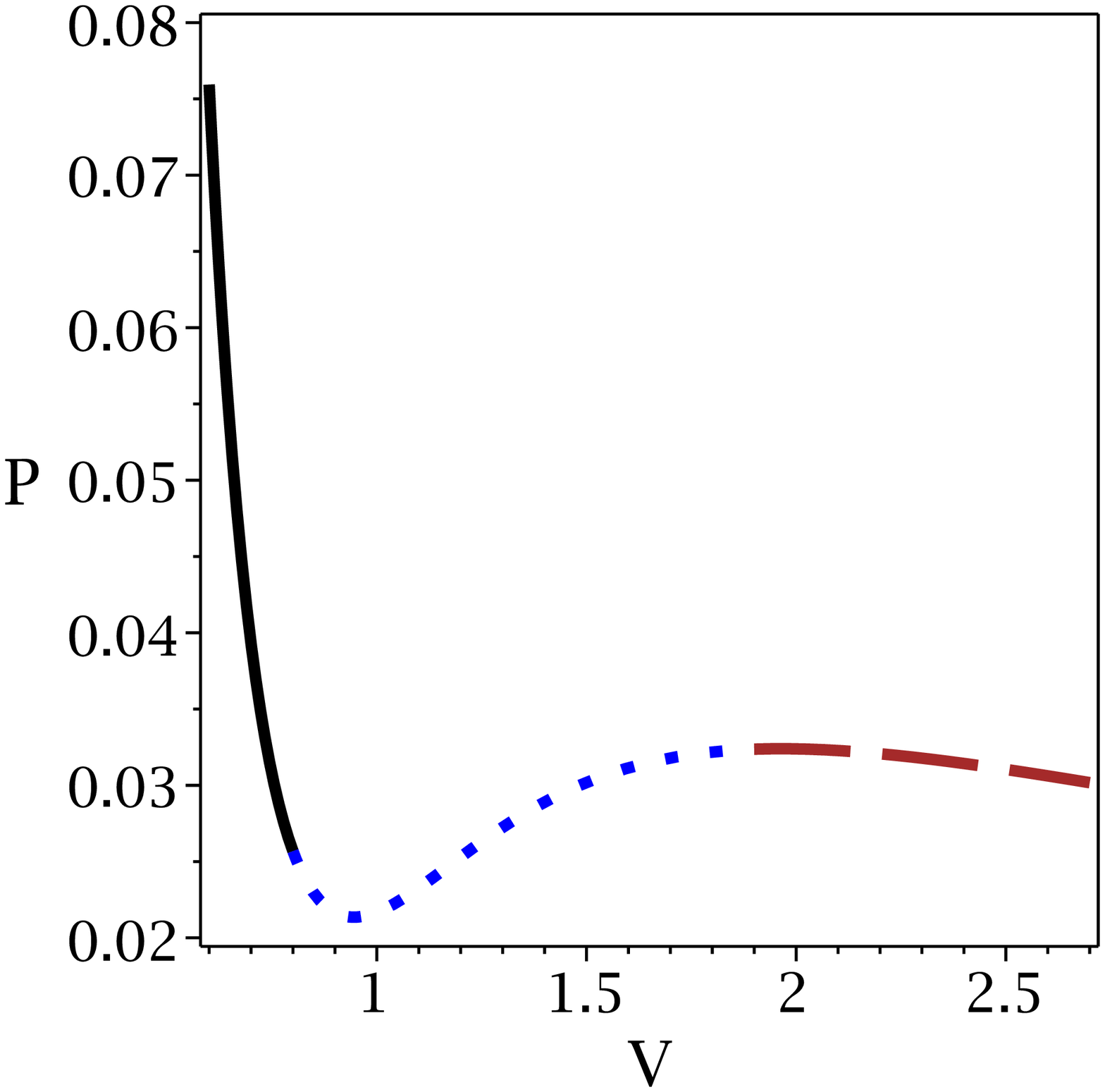} & \epsfxsize=7cm \epsffile{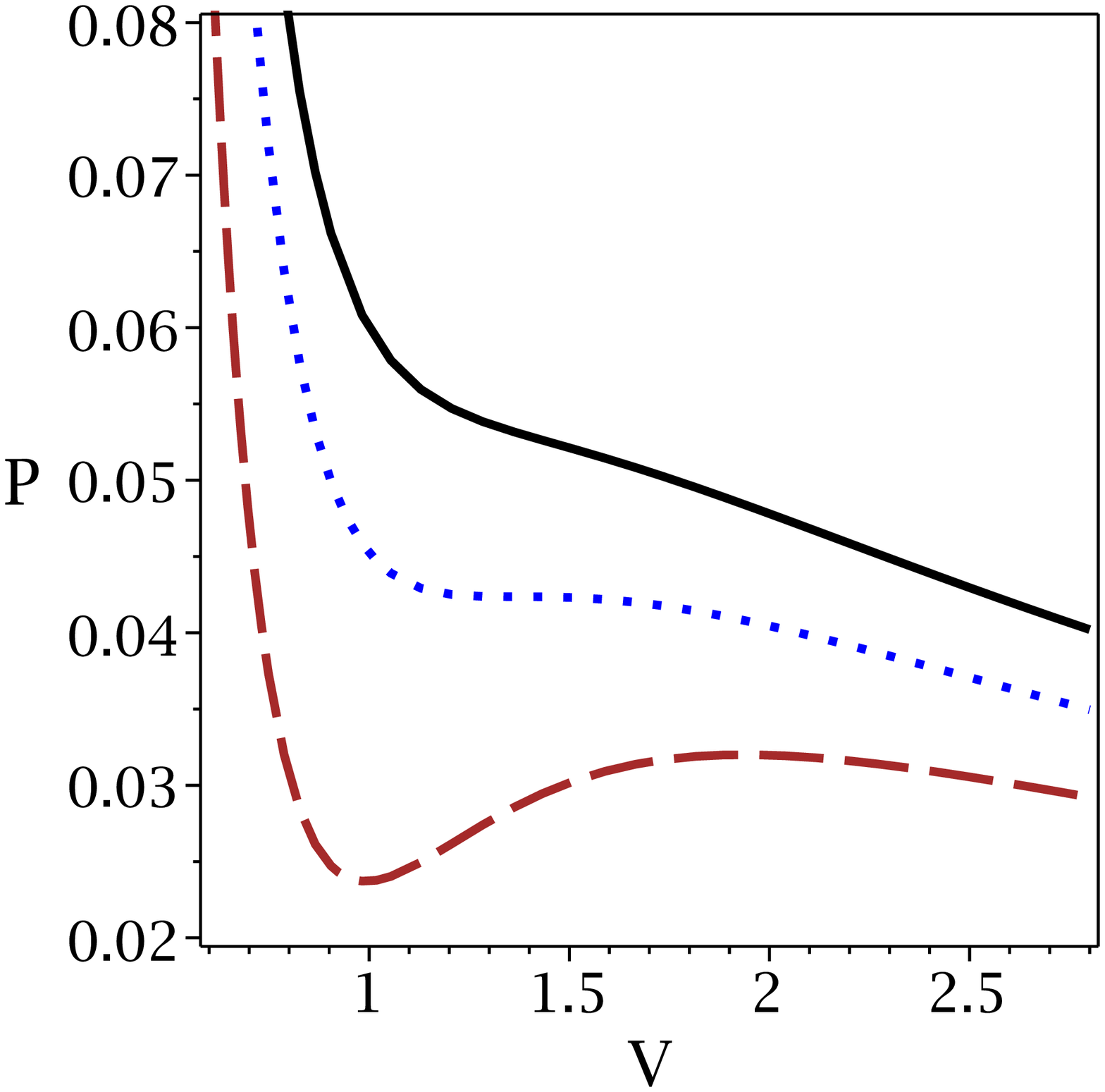}%
\end{array}
$%
\caption{\textbf{Left panel:} $P-V$ diagram for $T<T_{c}$. \newline
\textbf{Right panel:} $P-V$ diagrams for $T>T_{c}$ (continuous line), $%
T=T_{c}$ (doted line) and $T<T_{c}$(dashed line).}
\label{Fig4}
\end{figure}

In order to highlight the similarity between usual the heat engine and our
system under consideration, we plot $P-V$ diagrams for the above obtained
equation in Fig. (\ref{Fig4}), by considering fixed quantities for $m^{2}$
(the massive parameter), $c$, $c_{1}$, $c_{2}$ and $q$.

\begin{figure}[t]
\centering
$%
\begin{array}{c}
\epsfxsize=15cm \epsffile{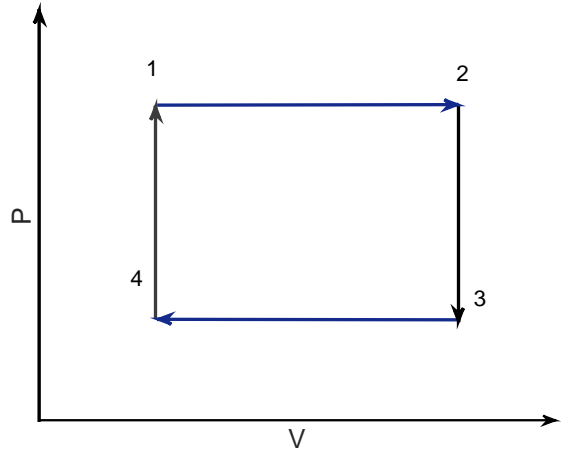}%
\end{array}
$%
\caption{$P$-$V$ diagram.}
\label{Fig5}
\end{figure}

Left panel of Fig. \ref{Fig4} shows that there are three areas for black
holes which encounter with a phase transition. The first area is related to
high pressure (continuous line in left panel in Fig. (\ref{Fig4})). The
black hole in this area has small radius and it is called small black hole
(SBH) region. The second area is related to unstable phase (dotted line in
left panel in Fig. (\ref{Fig4})). The third area is related to low pressure
case (dashed line in left panel of Fig. (\ref{Fig4})). In this area, black
holes have large radius, and it is known as large black hole (LBH) region.
On the other hand, for the temperatures more than critical temperature ($%
T>T_{C})$, the phase transition and the second area are disappeared (see
right panel of Fig. (\ref{Fig4}) for more details). In phase transition
point, the transition takes place between SBH to LBH. The opposite could
also take place in the case black holes horizon shrinking (the LBH to the
SBH). This enables us to define a classical cycle for black holes. It is
notable that the LBH lose $Q_{C}$ amount of heat along isothermal
contraction and the SBH absorbs $Q_{H}$ amount of heat along isothermal
expansion. Such transition also exists at high pressure. On the other hand,
an explicit expression for $C_{P}$ would suggest that there should be a new
engine which includes two isobars and two isochores/adiabatic similar to
Fig. (\ref{Fig5}). For this purpose, we can consider a rectangle cycle in
the $P-V$ plane in which this rectangle consists of two isobars ($%
1\rightarrow 2$ and $3\rightarrow 4$) and two isochores ($2\rightarrow 3$
and $4\rightarrow 1$), see Fig. (\ref{Fig5}) for more details. Also, a
possible scheme for this heat engine involves specifying values of
temperature where $T_{2}=T_{H}$ and $T_{4}=T_{C}$, in which $T_{H}$\ and $%
T_{C}$\ are temperatures of the warm and the cold reservoirs, respectively.
According to the fact that the paths of $1\rightarrow 2$ and $3\rightarrow 4$
are isobars, we find $P_{1}=P_{2}$ and $P_{3}=P_{4}$.

Now, we can calculated the work which is done in this cycle as
\begin{eqnarray}
W &=&\oint PdV \\
&&  \notag \\
W_{total} &=&W_{1\rightarrow 2}+W_{2\rightarrow 3}+W_{3\rightarrow
4}+W_{4\rightarrow 1}=W_{1\rightarrow 2}+W_{3\rightarrow 4}=P_{1}\left(
V_{2}-V_{1}\right) +P_{4}\left( V_{4}-V_{3}\right).
\end{eqnarray}

It is notable that the works which are done in the paths of $2\rightarrow 3$
and $4\rightarrow 1$ are isochores, therefore, these terms are zero. Using
Eqs. (\ref{TotalS}) and (\ref{V}), we have
\begin{equation}
W_{total}=\frac{4}{3\sqrt{\pi }}\left( P_{1}-P_{4}\right) \left(
S_{2}^{3/2}-S_{1}^{3/2}\right) .  \label{W}
\end{equation}

Also, the upper isobar (Fig. (\ref{Fig5})) will give the net inflow of heat
which is $Q_{H}$, so we have
\begin{equation}
Q_{H}=\int_{T_{1}}^{T_{2}}C_{P}\left( P_{1},T\right) dT.  \label{heat}
\end{equation}

Here, we want to obtain the efficiency of this cycle, so we use two
approximations:

$Case\ I:$ In limit of high pressure and considering temperature (Eq. (\ref%
{T1}) and the heat engine (Eq. (\ref{Cp})), we have
\begin{eqnarray}
T &\sim &2P\sqrt{\frac{S}{\pi }},  \label{TT} \\
&&  \notag \\
C_{P} &\sim &2S,  \label{Cpp}
\end{eqnarray}%
which by using Eq. (\ref{TT}), the entropy will be obtained as $S=\frac{\pi
T^{2}}{4P^{2}}$, and by replacing it in Eq. (\ref{Cpp}), we can find
\begin{equation}
C_{P}=\frac{\pi T^{2}}{2P^{2}}.
\end{equation}

This leads to obtain the $Q_{H}$ as
\begin{equation}
Q_{H}=\frac{\pi }{6P_{1}^{2}}\left( T_{2}^{3}-T_{1}^{3}\right) =\frac{4}{3%
\sqrt{\pi }}P_{1}\left( S_{2}^{3/2}-S_{1}^{3/2}\right) .  \label{QH11}
\end{equation}

Considering Eqs. (\ref{W}) and (\ref{QH11}), we can calculate the efficiency
of this cycle as
\begin{equation}
\eta =\frac{W}{Q_{H}}=1-\frac{P_{4}}{P_{1}},  \label{efficiency}
\end{equation}%
so we have
\begin{equation}
\eta =1-\frac{T_{C}}{T_{H}}.
\end{equation}

$Case\ II:$ In limit of high pressure and considering the path $1\rightarrow
2$ (remembering that in the path $1\rightarrow 2$, the entropy increases, we
can omit terms in which the entropy is in the denominator) in order to
obtain $Q_{H}$, the temperature will be
\begin{equation}
T=2P\sqrt{\frac{S}{\pi }}+\frac{m^{2}cc_{1}}{4\pi },  \label{Tem2}
\end{equation}%
which by using the above equation, we have
\begin{equation}
S=\frac{\pi T^{2}}{4P^{2}}+\frac{m^{2}cc_{1}T}{8P^{2}}+\frac{%
m^{4}c^{2}c_{1}^{2}}{64\pi P^{2}}.
\end{equation}

On the other hand, according the second approximation, we can use of the
first term of the heat capacity in which it will be the similar to Eq. (\ref%
{Cpp}). Using Eqs. (\ref{heat}) and (\ref{Cpp}), one can obtain $Q_{H}$ in
the following form
\begin{equation}
Q_{H}=\frac{\pi \left( T_{2}^{3}-T_{1}^{3}\right) }{6P^{2}}+\frac{%
m^{2}cc_{1}\left( T_{2}^{2}-T_{1}^{2}\right) }{8P^{2}}+\frac{%
m^{4}c^{2}c_{1}^{2}\left( T_{2}-T_{1}\right) }{32\pi P^{2}}.  \label{QH2}
\end{equation}

Now, we can investigate the efficiency of cycle by using Eqs. (\ref{W}) and (%
\ref{QH2}) which leads to
\begin{eqnarray}
\eta &=&\frac{W}{Q_{H}}=\left( 1-\frac{P_{4}}{P_{1}}\right) \left\{ 1-\frac{%
3m^{2}cc_{1}}{4\sqrt{\pi }P_{1}\left( \sqrt{S_{2}}-\sqrt{S_{1}}\right) }+%
\frac{3m^{4}c^{2}c_{1}^{2}}{8\pi P_{1}^{2}\left( S_{2}-S_{1}\right) }+%
\mathcal{O}\left( \frac{1}{P_{1}^{3}}\right) \right\} ,  \notag \\
&&  \notag \\
&=&\left( 1-\frac{T_{C}}{T_{H}}\right) \left\{ 1-\frac{3m^{2}cc_{1}}{4\sqrt{%
\pi }T_{H}\left( \sqrt{S_{2}}-\sqrt{S_{1}}\right) }+\frac{%
3m^{4}c^{2}c_{1}^{2}}{8\pi T_{H}^{2}\left( S_{2}-S_{1}\right) }+\mathcal{O}%
\left( \frac{1}{T_{H}^{3}}\right) \right\} .
\end{eqnarray}

Evidently, the massive parameter affects the efficiency of cycle for the
obtained black holes in massive gravity. It is notable that, when $m=0$, the
efficiency of cycle reduces to the efficiency of cycle for obtained black
holes in Einstein gravity.

\section{Exact efficiency formula}

Accordingly, if the thermodynamical cycle does not include the phase
transition area in the $P-V$ plane which leads to every temperature at the
fixed pressure only corresponds to one positive horizon radius $r_{+}$(or $V$%
) such that we can get the exact efficiency formula. In this case, we do not
have to calculate the approximate efficiency formula by appealing to high
temperature and high pressure limits. However when the phase transition
included in the cycle, the approximation method is necessary. For
simplicity, we just put the thermodynamical cycle in the third area, i.e.
stable large black hole region in Fig. \ref{Fig4} to avoid the phase
transition region, and we can obtain $Q_{H}$ as
\begin{equation}
Q_{H}=\int_{T_{1}}^{T_{2}}C_{P}(P_{1},T)dT=\int_{r_{1}}^{r_{2}}C_{P}(P_{1},T)%
\frac{\partial T}{\partial r}dr=Q_{H2}-Q_{H1},
\end{equation}%
where
\begin{gather}
Q_{H1}=\frac{6\pi ^{3/2}q^{2}+3cc_{1}m^{2}S_{1}^{3/2}+2\sqrt{\pi }%
S_{1}(3+3c^{2}c_{2}m^{2}+8P_{1}S_{1})}{12\pi \sqrt{S_{1}}}, \\
Q_{H2}=\frac{6\pi ^{3/2}q^{2}+3cc_{1}m^{2}S_{2}^{3/2}+2\sqrt{\pi }%
S_{2}(3+3c^{2}c_{2}m^{2}+8P_{1}S_{2})}{12\pi \sqrt{S_{2}}}.
\end{gather}

The efficiency is calculated as
\begin{equation}
\eta =\frac{W}{Q_{H}}=\left( 1-\frac{P_{4}}{P_{1}}\right) \times \frac{1}{1+%
\frac{3(\sqrt{S_{1}S_{2}}-\pi q^{2})}{8\sqrt{S_{1}S_{2}}(S_{1}+S_{2}+\sqrt{%
S_{1}S_{2}})P_{1}}+\frac{3cm^{2}(2cc_{2}\sqrt{\pi S_{2}}+c_{1}(S_{2}+\sqrt{%
S_{1}S_{2}}))}{16\sqrt{\pi S_{2}}(S_{1}+S_{2}+\sqrt{S_{1}S_{2}})P_{1}}}.
\end{equation}

Here, we focus on the large volume branch of solutions and therefore neglect
$q$ to leading order. This leads to
\begin{equation}
\eta =\left( 1-\frac{P_{4}}{P_{1}}\right) \left\{ 1-\frac{1}{P_{1}}\left(
\frac{3cm^{2}(2cc_{2}\sqrt{\pi S_{2}}+c_{1}(S_{2}+\sqrt{S_{1}S_{2}}))}{16%
\sqrt{\pi S_{2}}(S_{1}+S_{2}+\sqrt{S_{1}S_{2}})}\right) -\frac{3}{8P_{1}}%
\left( \frac{S_{2}^{\frac{1}{2}}-S_{1}^{\frac{1}{2}}}{S_{2}^{\frac{3}{2}%
}-S_{1}^{\frac{3}{2}}}\right) +O\left( \frac{1}{P_{1}^{2}}\right) \right\} .
\label{efficiency2}
\end{equation}

It is notable that when graviton mass is $m=0$, the efficiency calculated in
(\ref{efficiency2}) reduces to efficiency obtained in Ref. \cite{Johnson}.
The carnot efficiency is
\begin{equation}
\eta _{c}=1-\frac{T_{4}(S_{1},P_{4})}{T_{2}(S_{2},P_{1})}=1-\frac{\frac{1}{4%
\sqrt{\pi S_{1}}}\left[ 1+8S_{1}P_{4}-\frac{\pi q^{2}}{S_{1}}+m^{2}cc_{1}%
\sqrt{\frac{S_{1}}{\pi }}+m^{2}c^{2}c_{2}\right] }{\frac{1}{4\sqrt{\pi S_{2}}%
}\left[ 1+8S_{2}P_{1}-\frac{\pi q^{2}}{S_{2}}+m^{2}cc_{1}\sqrt{\frac{S_{2}}{%
\pi }}+m^{2}c^{2}c_{2}\right] },
\end{equation}%
where $T_{2}$ and $T_{4}$ refer to the highest and lowest temperatures in
the thermodynamical cycle, respectively. This fact holds if the
thermodynamical cycle is not putted in the second area (i.e. unstable black
hole phase) in Fig. (\ref{Fig4}). From the efficiency formula obtained
above, we can see that the graviton mass can affect the efficiency, but
whether the mass $m$ will improve the efficiency or reduce it, depends on
the choice of $c,c_{1}$ and $c_{2}$ parameters. In order to probe how these
parameters influence the efficiency, we plot some figures in the following
discussion.
\begin{figure}[tbph]
\centering
\includegraphics[height=2.4in,width=3in]{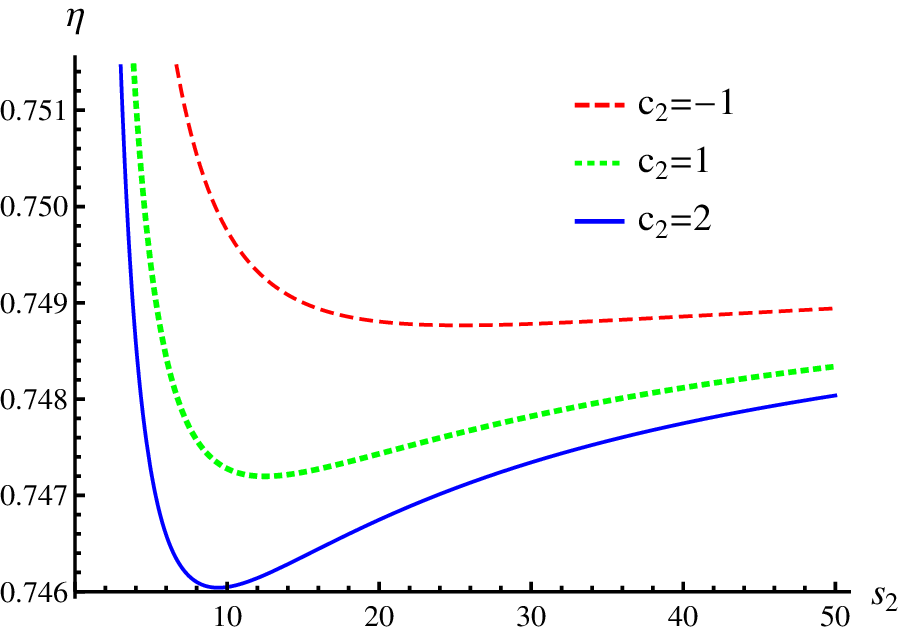}~~ %
\includegraphics[height=2.4in,width=3in]{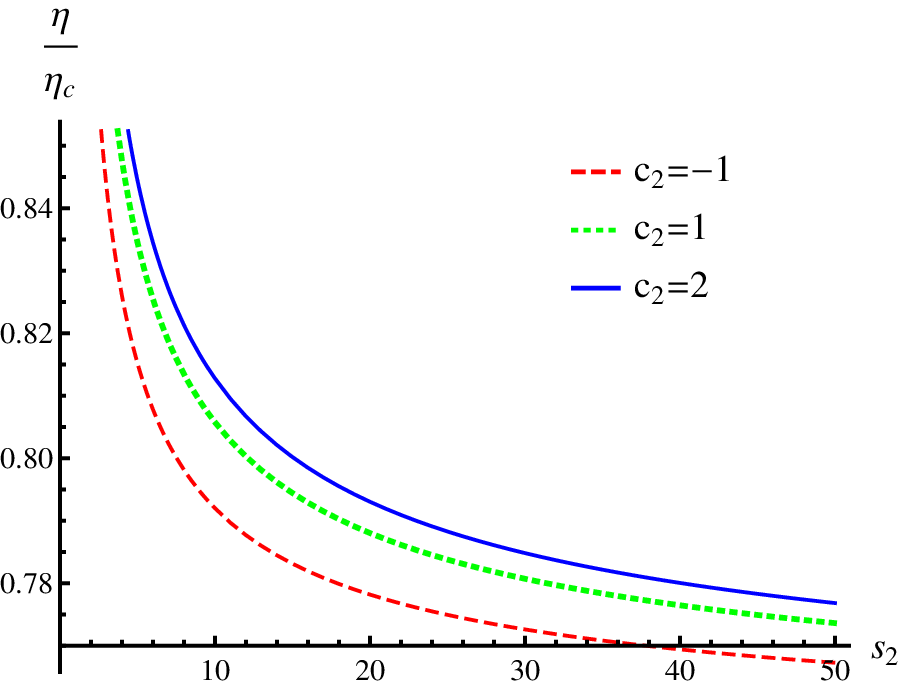}
\caption{The two figures above are plotted at 3 different parameters $c_{2}$
, here we take $P_{1}=4,P_{4}=1,S_{1}=1,q=1,c=c_{1}=1$ and graviton mass $%
m=1 $.}
\label{eff1}
\end{figure}
We plot $\eta $ and $\eta /\eta _{c}$ in Fig. (\ref{eff1}) under the change
of $S_{2}$ which shows that with the growth of $S_{2}$, the efficiency will
decrease to a minimal value and then monotonically increase to a maximum
value obtained by
\begin{equation}
\lim_{S_{2}\rightarrow \infty }\eta=\lim_{S_{2}\rightarrow \infty }\frac{\eta%
}{\eta_c} =1-\frac{P_{4}}{P_{1}} ,  \label{eta}
\end{equation}%
and increasing $c_{2}$ will lead to a lower efficiency, $\eta $. The
behavior of curves of $\eta $ is different from the figures plotted in Ref.
\cite{Liu1} which demonstrates that the efficiency influenced by
quintessence field would only monotonically decrease with the growth of $%
S_{2}$. From the right panel of Fig. (\ref{eff1}), we can see that the ratio
between efficiency $\eta $ and carnot efficiency $\eta _{c}$ will
monotonically decrease with the growth of $S_{2}$, and increasing $c_{2}$
parameter corresponds to a higher ratio, although for bigger $c_{2}$, the $%
\eta $ will be lower.
\begin{figure}[tbph]
\centering
\includegraphics[height=2.4in,width=3in]{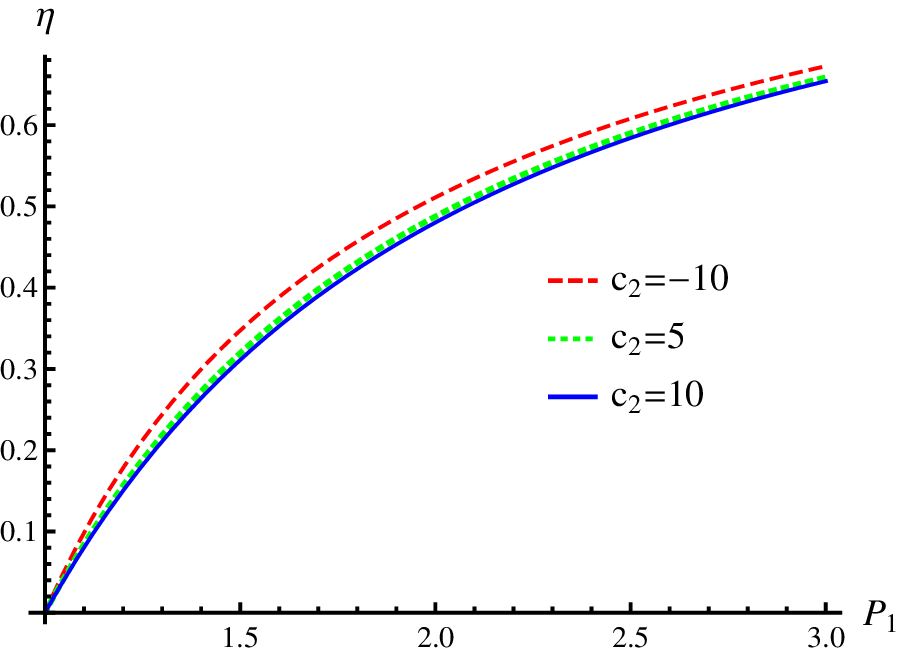}~~ %
\includegraphics[height=2.4in,width=3in]{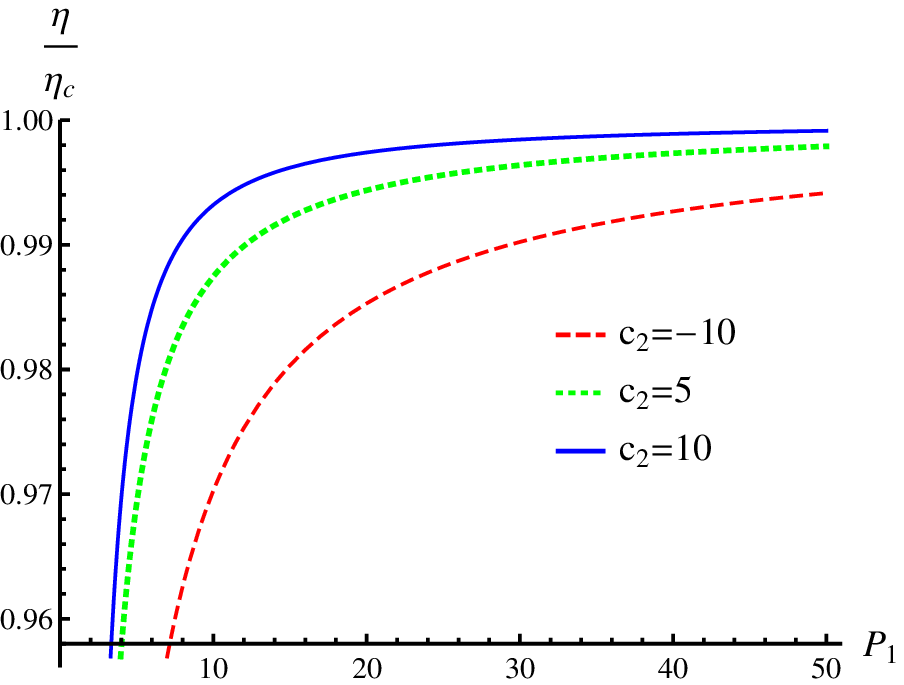}
\caption{The two figures above are plotted at 3 different parameters $c_{2}$
, here we take $P_{4}=1,S_{1}=8,S_{2}=12,q=1,c=c_{1}=1$ and graviton mass $%
m=1$.}
\label{eff2}
\end{figure}

In Fig. (\ref{eff2}), we can see that both the efficiency $\eta$ and $%
\eta/\eta_c$ will monotonously increase with the growth of pressure $P_1$,
and $\eta$ will infinitely approach the maximum efficiency, i.e. the Carnot
efficiency allowed by thermodynamics laws. In fact, in the high pressure
limit, we have
\begin{equation}
\lim_{P_1\rightarrow\infty} \eta=\lim_{P_1\rightarrow\infty}\frac{\eta}{%
\eta_c}=1.
\end{equation}

Previously, it was shown that it is possible to define van der Waals like
phase transition for non-spherical black hole solutions, and hence
horizon-flat and hyperbolic black holes \cite{PVMassIV}. Since the topology
of the horizon of black hole in massive gravity could be sphere, Ricci flat
or hyperbolic, corresponding to $k=1,0$ or $-1$, respectively, it would be
interesting to investigate the efficiency calculated on the horizon with
these three different topologies and then make a comparison. The temperature
for this topological black hole is \cite{PVMassIV}
\begin{equation}
T=\frac{1}{4\pi r_{+}}\left[ k-\Lambda r_{+}^{2}-\frac{q^{2}}{r_{+}^{2}}%
+m^{2}cc_{1}r_{+}+m^{2}c^{2}c_{2}\right] .  \label{Tem3}
\end{equation}

Based on Eq. (\ref{Tem3}), we can get the efficiency
\begin{equation}
\eta _{k}=\frac{W}{Q_{H}}=\left( 1-\frac{P_{4}}{P_{1}}\right) \times \frac{1%
}{1+\frac{3(k\sqrt{S_{1}S_{2}}-\pi q^{2})}{8\sqrt{S_{1}S_{2}}(S_{1}+S_{2}+%
\sqrt{S_{1}S_{2}})P_{1}} + \frac{3cm^{2}(2cc_{2}\sqrt{\pi S_{2}}+c_{1}(S_{2}+%
\sqrt{S_{1}S_{2}}))}{16\sqrt{\pi S_{2}}(S_{1}+S_{2}+\sqrt{S_{1}S_{2}})P_{1}}}%
.
\end{equation}

It is obvious that when all the other variables are fixed, we have
\begin{equation}
\eta _{-1}>\eta _{0}>\eta _{1},
\end{equation}
which means that the efficiency of black hole engines in massive gravity
with hyperbolic horizon is higher than that of black holes with flat
horizon. In addition, one finds the spherical black holes have the lowest
efficiency.

\section{Conclusions}

In this paper, we have considered charged black holes in the presence of
massive gravity. It was shown that the temperature and pressure of these
black holes are functions of the massive gravity. This indicated the
dependency of thermal phase transition points on the massive gravity and its
parameters as well.

Next, by using the phase transition points, it was shown that it is possible
to define a cyclic thermodynamical behavior consisting two isobars and two
isochores. This cycle could be interpreted as a heat engine. In other words,
it was possible to show that charged black holes in the massive gravity
admits a heat engine. Interestingly, it was shown that contrary to other
cases, it is possible to define heat engine for non-spherical black holes as
well. In other words, due to contributions of the massive gravity, the heat
engine could be constructed for horizon flat and hyperbolic black holes and
it is not limited only to spherical black holes.

The expressions extracted for efficiency and heat were shown to be massive
gravity dependent. The effects of massive gravity were highly dependent on
the sign and values of the massive coefficients, $c_{i}$'s. Especially for $%
c_{2}$, it was shown that efficiency is a decreasing function of this
parameter. If we consider only positive and non-zero values of the $c_{i}$,
one can conclude that efficiency of heat engine in the massive gravity is
smaller comparing to the absence of the massive gravity. This highlights the
contributions of massive gravity on the properties of heat engine.

To have a better understanding of the massive gravity's impacts on the heat
engine efficiency, we plotted Figs. (\ref{eff1}) and (\ref{eff2}) of the
efficiency under the change of entropy $S_2$ (or larger black hole with
thermodynamical volume $V_2$) and pressure $P_1$ at the fixed parameters $%
c_i $'s which can be chosen to promote or reduce efficiency, respectively.
As it is shown by Fig. (\ref{eff1}), under the chosen parameters $c_i$, we
find that with the grow of volume difference $\Delta V=V_2-V_1$/or $\Delta
S=S_2-S_1$ between the smaller black hole with thermodynamical volume $V_1$
(or $S_1$) and the larger black hole with volume $V_2$ (or $S_2$), the
efficiency $\eta$ of the thermodynamics cycle will decrease to a minimal
value at first and then gradually increase from the minimal $\eta$ to the
limit of $\eta=1-\frac{P_4}{P_1}$ when $\Delta V$ (or $\Delta S$) goes to
infinity. For the ratio $\frac{\eta}{\eta_c}$, it will monotonously decrease
with the grow of $\Delta V$ and it is interesting to note that the limits of
$\frac{\eta}{\eta_c}$ is equal to the limits of $\eta$ which depends on the
pressures $P_1$ and $P_4$ as it was shown by Eq. (\ref{eta}). On the other
hand, the efficiency $\eta$ is lower but the ratio $\frac{\eta}{\eta_c}$ is
higher when the thermodynamic cycle has a bigger $c_2$. Furthermore, Fig. (%
\ref{eff2}) showed that for the bigger pressure difference $\Delta P=P_1-P_4$%
, the efficiency $\eta$ and ratio $\frac{\eta}{\eta_c}$ will be larger. When
$\Delta P\rightarrow\infty$, $\eta$ and $\frac{\eta}{\eta_c}$ will
infinitely approach value $1$ which is not allowed to be exceeded.

One of the possibilities provided for black holes in the presence of massive
gravity is existence of van der Waals like behavior for non-spherical black
holes. Such possibility was not reported for other black holes in the
presence of different matter fields and gravities. Using this possibility,
we were able to have a heat engine for non-spherical black holes which was
not observed before. We also conducted a study regarding the effects of
topological structure of the black holes on efficiency of the heat engine.
Interestingly, it was shown that the smallest efficiency for heat engine
belongs to spherical black holes while the highest one was provided for
hyperbolic black holes.

At last but not least, according to the AdS/CFT correspondence, it would
also be of interest to have a deeper holographic understanding of the black
hole heat engines in AdS space. It has been argued that such heat engines
may have interesting holographic implications because the engine cycle
represents a journal through a family of holographically dual large $%
\mathcal{N}$ field theories as explained in Ref. \cite{JohnsonI}. On the
other hand, considering that the energy of the black hole can be extracted
by the way of transferring heat to mechanical work, the black hole heat
engine may be regarded as a possible energy source for the high energy
astrophysical phenomena near the black holes. We should point out that the
topological black holes as heat engines in higher dimensional massive
gravity are also worthy to be investigated, which we leave this issue for
future work.

\begin{acknowledgements}
We thank Shiraz University Research Council. This work has been
supported financially by the Research Institute for Astronomy and
Astrophysics of Maragha, Iran.
\end{acknowledgements}

\end{document}